\documentclass[twocolumn,showpacs,preprintnumbers,amsmath,amssymb]{revtex4}

\newcommand{\Med}[1]{\left\langle #1 \right\rangle}
\renewcommand{\H} {{\cal H}}
\newcommand{\D} {{\cal D}}
\newcommand{\qs} {{\tilde q}}

\usepackage[dvips]{graphicx}

\begin{document}
\title{Numerical estimates of the finite size corrections to the free 
energy of the SK model using Guerra--Toninelli interpolation}
\author{Alain Billoire}
\date{\today}
\affiliation{
  Service de physique th\'eorique,
  CEA Saclay,
  91191 Gif-sur-Yvette, France.
}

\begin{abstract}
I use an interpolation formula, introduced recently by Guerra and
Toninelli in order to prove the existence of the free energy of the
Sherrington--Kirkpatrick spin glass model in the infinite volume limit, to
investigate numerically the finite size corrections to the free energy
of this model. The results are compatible with a $(1/12 N) \ln(N/N_0)$
behavior at $T_c$, as predicted by Parisi, Ritort and Slanina, and a
$1/N^{2/3}$ behavior below $T_c$.
\end{abstract}

\pacs{PACS numbers: 75.50.Lk, 75.10.Nr, 75.40.Gb}

\maketitle

Many years after their experimental discovery, spin glasses remain a
challenge for experimentalists, theoreticians and more recently
computer scientists and mathematicians. Numerical simulations have
been used heavily in order to investigate their physical
properties. Numerical simulations are obviously limited to finite
systems.  Simulations of spin glasses are indeed limited to very small
systems, due to the need to repeat the simulation for many disorder
samples (this is related, at least for mean field models, to the lack
of self-averaging), and to the bad behavior, as the system size grows,
of all known algorithms. A detailed understanding of finite size
effects of spin glass models is accordingly highly desirable. The
problem is also interesting in its own sake\cite{PRS1,PRS2,Moore}.

Here I study the finite size behavior of the Sherrington--Kirkpatrick
model\cite{SK} (SK model), a well know infinite connectivity model,
introduced originally in order to have a solvable starting point for
the study of ``real" finite connectivity spin glasses, and that turned
out to have a complex fascinating structure, to the point of
becoming\cite{Tala} ``a challenge for mathematicians".

The partition function of the $N$ sites Sherrington--Kirkpatrick model is

\begin{eqnarray}
\nonumber
Z_N &=&\exp(-\frac{N f_N(T)}{T})\\ 
\nonumber
&=& \sum_{\{\sigma\}}
\exp(\frac{1}{\sqrt{N} T}\sum_{1\leq i< j\leq N}J_{i,j} \sigma_i
\sigma_j),
\end{eqnarray}

where $T$ is the temperature, the $\sigma_i$'s are Ising spins, and
the $J_{i,j}$'s independent, identically distributed, Gaussian random
numbers with zero mean and unit square deviation.  In the paramagnetic
phase, the finite size behavior of the disorder-averaged free energy
$\overline f_N(T)$ can be computed, using the replica method, as an
expansion in powers of $1/N$, as shown by Parisi, Ritort and
Slanina\cite{PRS1}. One starts from the equation~\cite{BM1}

\begin{eqnarray}
\frac{\overline{f_N(T)}}{T} &=& -\ln 2-\frac{1}{4T^2}\\
\nonumber
&-&\lim_{n\to 0}\frac{1}{n N}
\ln \int(\prod_{a<b} \sqrt{\frac{NT^2}{2\pi}} d\qs_{a,b}) e^{-N \H(\qs)},
\end{eqnarray}

\begin{eqnarray}
\H(\qs)&=&
\frac{\tau}{4}\sum \qs_{a,b}^2
-\frac{1}{6}\sum \qs_{a,b}\qs_{b,c}\qs_{c,a}\\
\nonumber
&-&\frac{1}{8}\sum \qs_{a,b}\qs_{b,c}\qs_{c,d}\qs_{d,a}
+\frac{1}{4}\sum \qs_{a,b}^2 \qs_{a,c}^2\\
\nonumber
&-&\frac{1}{12}\sum \qs_{a,b}^4,
\end{eqnarray}

where $\tau=(T^2-1)/2$, the field $\qs$ is a real symmetric $n\times
n$ matrix, with $\qs_{a,a}=0$. The matrix $\qs$ has been rescaled by a
factor $1/T^2$ (namely $\qs = q/T^2$), and the terms of order $\qs^5$
and higher have been omitted from the effective Hamiltonian $\H(\qs)$.
In the paramagnetic phase, on can expand the integrand around the
saddle point $\qs_{a,b}=0$. Keeping the quadratic term only in $\H$,
one obtains

\begin{equation}
\frac{\overline{f_N(T>1)}}{T}=-\ln 2-\frac{1}{4T^2}
-\frac{1}{4 N}\ln(2\tau/T^2).
\end{equation}

Treating perturbatively the interaction terms in $\H$ one
builds\cite{PRS1} a loop expansion for the finite size corrections to
the free energy.  The $k$ loops  term goes like $1/N^k$, with the
most diverging contribution as $T\to 1$ (namely $\propto 1/(N^k
\tau^{3(k-1)})$) coming from the order $\qs^3$ term in the
Hamiltonian. Summing up these contributions, one obtains~\cite{PRS1}
at the critical temperature

\begin{equation}
\frac {\overline{f_N(T=1)}}{T}=-\ln 2-\frac{1}{4}
+\frac{\ln N}{12 N}+\frac{f_{(-1)}}{N}+\cdots
\label{FTC}
\end{equation}

The computation~\cite{PRS1} of the constant $f_{(-1)}$ requires a
non-perturbative extrapolation.  The various prescriptions tried for
this extrapolation gave unfortunately quite different values for
$f_{(-1)}$, in the approximate range $[-0.2,+0.2]$.

It is not known how to extend the above analysis to the spin-glass
phase below $T_c$.  Numerical works indicate that the ground state
energy (or zero temperatures internal energy) scales
like\cite{Palassini1,Palassini2, BKM, Boettcher,KKLJH}
$\overline{e_N}-\overline{e_{\infty}}\propto N^{-2/3}$ (this result is
exact for the spherical SK model\cite{Andreanov}), like the internal
energy at $T_c$\cite{PRS1}.

In this brief report, I introduce a numerical method to compute the
finite size corrections to the free energy of the
Sherrington--Kirkpatrick model, based on Guerra and Toninelli
interpolation method.  Guerra and Toninelli
introduced the partition function

\begin{eqnarray}
\nonumber
Z_N(t)&=&\sum_{\{\sigma\}}\exp\Biggl(\frac{1}{T}\Bigl(\sqrt{\frac tN}\sum_{1\le i<j\le N}
J_{ij}\sigma_i\sigma_j\\
&+&\sqrt{\frac {1-t}{N_1}}\sum_{1\le i<j\le N_1}
J'_{ij}\sigma_i\sigma_j\\\nonumber
&+&\sqrt{\frac {1-t}{N_2}}\sum_{N_1 <i<j\le N}
J''_{ij}\sigma_i\sigma_j\Bigr)\Biggr),
\end{eqnarray}

that involves a parameter $t$ that interpolates between the SK model
with $N$ sites ($t=1$) and a system of two uncoupled SK models with
$N_1$ and $N_2=N-N_1$ sites ($t=0$). In what follows $N_1=N_2=N/2$.
The $J$s, $J'$s and $J''$s are independent identically distributed
Gaussian random numbers.  It is easy to show that

\begin{eqnarray}
\nonumber
\frac{f_{N}-f_{N/2}}{T}&=&\frac{1}{4 T^2}\int_0^1 dt 
\overline{
\Med{(q_{12})^2-\frac{1}{2}(q_{12}^{(1)})^2-
\frac{1}{2}(q_{12}^{(2)})^2}}
\\
&=&\frac{1}{4T^2}\int_0^1 dt\ {\D}(t)\qquad \D(t) \geq 0,
\label{DeltaF}
\end{eqnarray}

where

\begin{eqnarray}
q_{12}&=&\frac{1}{N} \sum_{i=1}^N \sigma_i\tau_i, \quad
q_{12}^{(1)}=\frac{2}{N} \sum_{i=1}^{N_1} \sigma_i\tau_i, \\
\nonumber
q_{12}^{(2)}&=&\frac{2}{N} \sum_{i=N_1+1}^{N} \sigma_i\tau_i.
\end{eqnarray}

\begin{figure}[tbh]
  \centering
  \includegraphics[width=0.350\textwidth,angle=270]{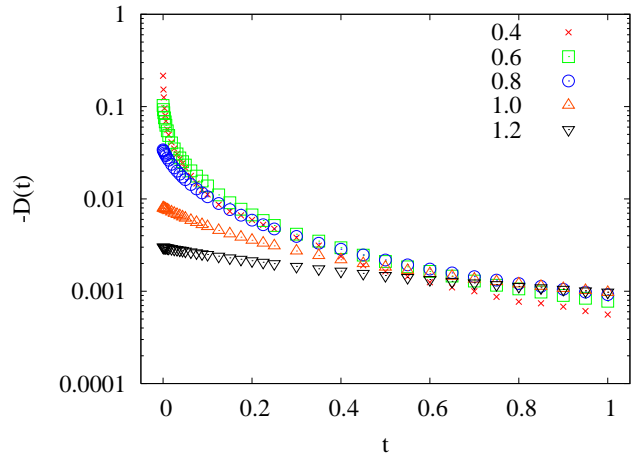}
  \caption{(Color online) Minus $\D(t)$ as a function of the
  interpolation parameter $t$ (both in logarithmic scale) for $N=1024$
  and temperatures $0.4, 0.6, \ldots, 1.2$.}
\label{figure1}
\end{figure}

The right hand side of equation~\ref{DeltaF} can be evaluated with
Monte Carlo simulation.  I use the Parallel Tempering algorithm, with
$T\in [0.4,1.3]$ and uniform $\Delta T=0.025$.  A total of $2\ 10^5$
sweeps of the algorithm was used for every disorder sample.  The
quenched couplings have a binary distribution in order to speed up the
computer program (as shown in reference~\cite{PRS1}, the leading
finite size correction is the same for the binary and Gaussian
couplings).  Systems of sizes $N$ from $128$ to $1024$ have been
simulated with $128$ disorder samples for each system size (but for
$N=1024$, where I used $196$ samples).  The integration over $t$ was
done with the trapezoidal rule, with $39$ non uniformly spaced
points. Integrating with only half of the points makes a very small
effect on the integrand (smaller than the estimated statistical
error).

Figure~\ref{figure1} shows the integrand $D(t)$ as a function of $t$
for the largest system and several temperatures.  The integrand is
concentrated around $t=0$, and I have chosen the discretization of $t$
accordingly.  One notices that $\D(t=0)$ is more and more negative as
$T$ decreases, as predicted by the formula
$\D(t=0)=-\overline{<(q_{12})^{2}>}$, and that $\D(t=1)$ is weakly
dependent on $T$, as expected from the identity $\D (t=1) =
{1}/(N-1)(\overline{<(q_{12})^2>}-1)$, which is weakly dependent on
$T$ (for not too small T's) since $\overline{<(q_{12})^2>}$ is small
compared to one.

\begin{figure}[tbh]
\centering
\includegraphics[width=0.350\textwidth,angle=270]{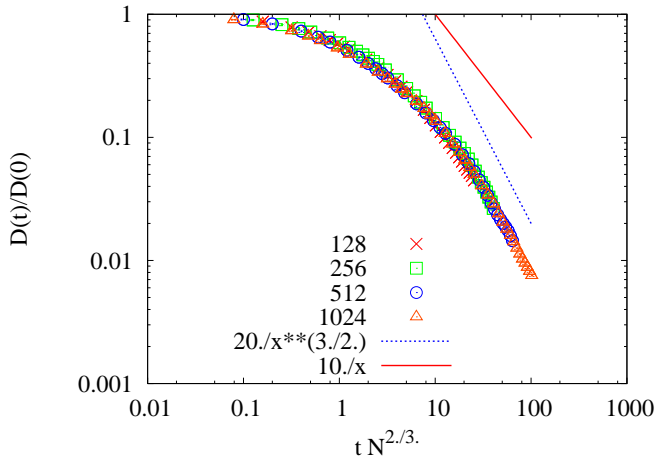}
  \caption{(Color online) ${\D(t)}/{D(t=0)}$ as a function of $t
  N^{2/3}$ (both in logarithmic scale), for $T=0.6$.  The orange line
  (full line) shows the $1/x$ behavior, the blue line (dotted line)
  shows the $1/x^{3/2}$ behavior. Clearly $D(t)$ grows faster than
  $1/x$ for large $x$. The precise behavior of $\D(t)$ is not
  essential for my argument, as soon as it decays faster than $1/x$.}
\label{figure2}
\end{figure}

In the low $T$ phase, a remarkable scaling is observed if one plots
the ratio ${\D(t)}/{D(t=0)}$ as a function of $t N^{2/3}$, as shown in
figures~\ref{figure2}.  It means that, to a good approximation, one
has $\D(t)/\D(0)=F(t N^{2/3})$, with $F(x)$ decaying faster than $1/x$
for large $x$, making the integral in equation~\ref{DeltaF} converge.
One has accordingly in the low $T$ phase $f_N-f_{\infty}\propto
1/N^{2/3}$.  A temperature independent exponent $2/3$ for the free
energy is in contradiction with the claims of~\cite{KC} that the
internal energy scales like $e_N-e_{\infty}\propto 1/N^{x(T)}$, with
an exponent $x(T)$ that is compatible with $2/3$ for both $T=0$ and
$T_c$ but reaches a minimum $\approx 0.54$ between. The results of
reference\cite{KC} are based however on Monte Carlo simulations of
relatively small systems with $N$ up to $196$.  Analyzing the data for
the internal energy produced during the simulation of
reference\cite{BM}, witch include systems with up to $4096$ spins, one
finds\cite{moi} an exponent that is much closer to $2/3$, with
deviations that are presumably explained by the proximity of the
critical point and by the very slow convergence of the expansion of
$e_N-e_{\infty}$ in inverse powers of $1/N$ (at $T_c$, the expansion
parameter is\cite{PRS1} $1/N^{1/3}$).

\begin{figure}[tbh]
  \centering
  \includegraphics[width=0.350\textwidth,angle=270]{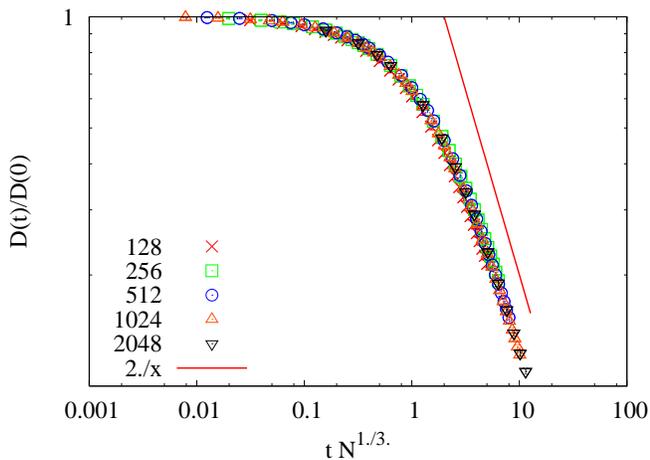}
  \caption{(Color online) ${\D(t)}/{D(t=0)}$ as a function of $t
  N^{1/3}$ (both in logarithmic scale), for $T=T_c$.  The orange line
  (straight line) shows the expected $1/x$ behavior, in order to guide
  the eyes.}
\label{figure4}
\end{figure}

The situation is different at $T_c$, as shown in figure~\ref{figure4},
the ratio ${\D(t)}/{D(t=0)}$ scales with a different exponent, namely
like $F(t N^{1/3})$, with a large $x$ behavior compatible with
$F(x)\propto 1/x$ (Although much larger system sizes would be needed
in order to be sure that the system really approaches this asymptotic
behavior). This is in agreement with formula~\ref{FTC} (In this model
one has $\beta/\nu=2$).  The data presented at $T_c$
(Figures~\ref{figure4} and~\ref{figure6}) include the results of an
additional simulation of a system with $N=2048$ sites, limited to the
(cheap to simulate) paramagnetic phase, with $T\in [1.0,1.3]$, $\Delta
T=0.025$, with 128 disorder samples, and a $15$ points discretization
of $t$.

\begin{figure}[tbh]
  \centering
  \includegraphics[width=0.350\textwidth,angle=270]{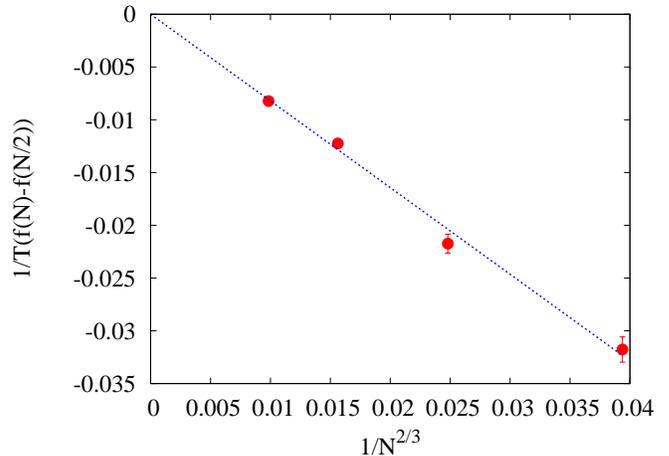}
  \caption{(Color online) Numerical data for $(f_N-f_{N/2})/T$ as a
  function of $1/N^{2/3}$, together with a numerical fit to the data of
  the form $(f_N-f_{N/2})/T=-A/N^{2/3}$, with $A=0.82\pm 0.02$ (blue
  dotted line).  Here $T=0.4$, $N=128, 256, 512\ {\rm and}\ 1024$.}
\label{figure5}
\end{figure}

Figure~\ref{figure5} shows, as a function of $1/N^{2/3}$, my
estimates, after integrating numerically equation~\ref{DeltaF}, of
$(f_N-f_{N/2})/T$ at $T=0.4$,  compared to the result of a linear fit
$(f_N-f_{N/2})/T= - A/N^{-2/3}$, with $A= 0.82\pm 0.02$ and
$\chi^2=4.9$.  The agreement is good within estimated statistical
errors.  A similar agreement is obtained for other values of $T$ in
the spin glass phase (e.g.  $A=0.39 \pm 0.01$ with $\chi^2=3.6$ for
$T=0.6$, and $A=0.18 \pm 0.01$ with $\chi^2=33$ - a large value
presumably related to the proximity of the critical point - for
$T=0.8$).  Figure~\ref{figure6} shows my estimates for
$(f_N-f_{N/2})/T$ at $T_c$ as a function of $1/N$, together with the
prediction of equation~\ref{FTC}.  A good agreement (with $\chi^2=4.3$
if one excludes the $N=128$ data from the fit) is obtained using the
value $1/N_0=7.8\pm 0.2$, namely $f_{(-1)}=\ln(7.8)/12=0.17\ldots$,
within the range of results presented by Parisi, Ritort and Slanina
\cite{PRS1}.

\begin{figure}[tbh]
  \centering
  \includegraphics[width=0.350\textwidth,angle=270]{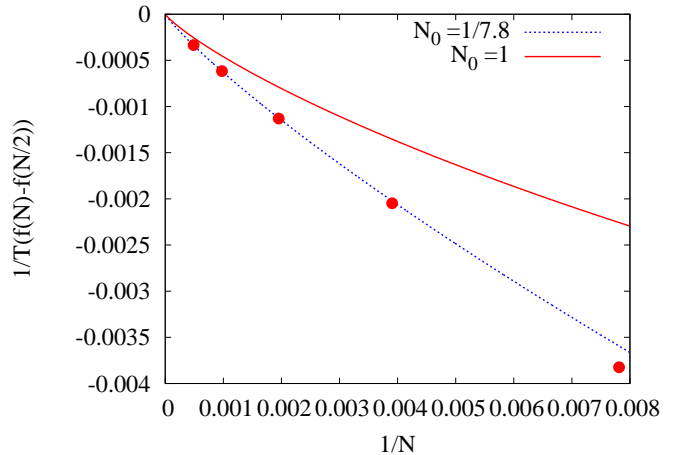}
  \caption{(Color online) Numerical data for $(f_N-f_{N/2})/T$ as a
  function of $1/N$, together with the behavior implied by the
  equation: $f_N/T=f_{\infty}/T+1/(12 N)\ln{N/N_0}$. The orange line
  (full line) is drawn with the value $N_0=1$. The blue line (dotted
  line) is drawn with the value $1/7.8$, from a fit to the data.  Here
  $T=1$, $N=128, 256, \ldots,2048$.}
\label{figure6}
\end{figure}

In conclusion, I have shown that the Guerra--Toninelli interpolation
provides an efficient method to evaluate numerically the finite size
corrections to the free energy of the Sherrington--Kirkpatrick model.
The integrand $\D(t)$ exhibits a remarkable scaling as a function of the
interpolation parameter $t$ and system size $N$.  At the critical
temperature, the results for the free energy are in agreement with the
predicted $(1/12 N) \ln(N/N_0)$ leading behavior of the finite size
corrections, and give the estimate $N_0\approx 1/7.8$.  In the low
temperature phase, the results indicate that the leading corrections
behave like $N^{-2/3}$ for both the internal energy and the free
energy of the model.

\section{Acknowledgments}
This work originates from a discussion with Giorgio Parisi. I thank
him for his enlightening advises. I also acknowledge email exchanges
with Frantisek Slanina, Andrea Crisanti, Ian Campbell and Helmut
Katzgraber. The simulation was performed at CCRT, the CEA computer
center at Bruy\`eres-le-Ch\^atel, using $\approx 32000$ hours of $1.25
$ GHz alpha EV68.

\end{document}